\documentclass[pra,
                twocolumn,
                preprintnumbers,
                amsmath,
                amssymb,
                showpacs,
                superscriptaddress]{revtex4-1}
\usepackage{graphicx} 
\usepackage{dcolumn}  
\usepackage{bm}       
\usepackage{amsfonts}
\usepackage{mathtools}
\usepackage[colorlinks=true,linkcolor=blue,citecolor=red,urlcolor=blue]{hyperref}
\usepackage{color}
\usepackage{natbib}
\newcommand{\Eion}{\mbox{$E_{\mbox{\scriptsize ion}}$}}

\newcommand{\vnor}{\mbox{$v_{\mbox{\scriptsize nor}}$}}
\newcommand{\vpar}{\mbox{$v_{\mbox{\scriptsize par}}$}}

\newcommand{\eq}[1]{Eq.~(\ref{#1})}
\begin{document}

\title{Resonant charge-transfer in grazing collisions of H$^-$ with vicinal nanosurfaces on Cu(111), Au(100) and Pd(111) substrates: A comparative study}%

\author{John Shaw}
\email[]{jshaw1969@gmail.com}
\affiliation{%
Department of Natural Sciences, D.L.\ Hubbard Center for Innovation,
Northwest Missouri State University, Maryville, Missouri 64468, USA}

\author{David Monismith} 
\affiliation{%
Software Maintenance Group, Tinker AFB, Oklahoma, USA}

\author{Yixiao Zhang}
\affiliation{%
Department of Natural Sciences, D.L.\ Hubbard Center for Innovation,
Northwest Missouri State University, Maryville, Missouri 64468, USA}
\altaffiliation{Cornell University, New York, USA}

\author{Danielle Doerr} 
\affiliation{%
Department of Natural Sciences, D.L.\ Hubbard Center for Innovation,
Northwest Missouri State University, Maryville, Missouri 64468, USA}
\altaffiliation{}

\author{Himadri S. Chakraborty}
\email[]{himadri@nwmissouri.edu}
\affiliation{%
Department of Natural Sciences, D.L.\ Hubbard Center for Innovation,
Northwest Missouri State University, Maryville, Missouri 64468, USA}

\date{\today}



\begin{abstract}
We compare the electron dynamics at monocrystalline Cu(111), Au(100) and Pd(111) precursor substrates with vicinal nanosteps. The unoccupied bands of a surface superlattice are populated \textit{via} the resonant charge transfer (RCT) between the surface and a H$^-$ ion that flies by at grazing angles. A quantum mechanical wave packet propagation approach is utilized to simulate the motion of the active electron where time-evolved wave packet densities are used to visualize the dynamics through the superlattice. The survived ion fraction in the reflected beam generally exhibits modulations as a function of the vicinal terrace size and shows peaks at those energies that access the image state subband dispersions. However, differences in magnitudes of the ion-survival as a function of the particular substrate selection as well as the ion-surface interaction time based on the choice of two ion-trajectories are examined. A square well model producing standing waves between the steps on the surface explains well the energies of the maxima in the ion survival probability for all the metals considered, indicating that the primary process of confinement induced subband formation is rather robust. The work may motivate measurements and applications of shallow-angle ion-scattering spectroscopy to access electronic substructures in periodically nanostructured surfaces.
\end{abstract}
\maketitle

\section{Introduction}

Vicinal surfaces are the simplest prototypes of lateral nanostructures on the surface which are thought to closely mimic rough regions of industrial surfaces. A vicinal surface is obtained by cutting a monocrystalline surface along a direction that somewhat deviates from a major crystallographic axis. These repeated ``miscuts" followed by subsequent recoveries form regular and uniform arrays of linear steps that can be polished by suitable ultra-high vacuum methods. Such high Miller index surfaces can be critical for their catalytic properties, without losing the lattice periodicity~\cite{pratt}. The electronic motions in these surfaces can be of particular fascination because of the possibility of the scattering of electrons at step edges to induce confinements resulting in subband dispersions. Vicinal steps provide nano-pockets to nucleate low-dimensional structure where the inclination angle can tune structure-substrate coupling, which is important in controlling their chemical properties~\cite{tegencamp09}. Also, photoelectron spectroscopy measured Ag nanostripes on step edges of the Cu vicinal to induce surface state splitting into bimetallic subbands from step-scattering and size quantization~\cite{schiller05}. Therefore, to better understand the dynamics and transport phenomenon in such functionalized and designed vicinals, knowledge of electron dispersions of the naked vicinal is important.  

Several metallic vicinal surfaces are of particular interest due to the presence of a Shockley surface state on and a Rydberg series of image states above the corresponding flat surface. These states arise from a broken translational symmetry along the direction perpendicular to the surface which localizes them in this direction by a projected band gap. Therefore, the electronic dispersions of these surface and image states should modify by the vicinal nano-stepping \textit{via} the superlattice effects. Alterations of the electronic properties of surface electrons have been detected for Cu and Au vicinals by scanning tunneling microscopy in the real space~\cite{didiot,suzuki,bolz} as well as by angle-resolved photoemissions in momentum space with synchrotron radiation~\cite{mugarza,mugarza01,ortega}.  Ultraviolet photoelectron spectroscopy has indicated unique two-dimensional Shockley surface states on (332) and (221) vicinals of Cu~\cite{baumberger,greber}. Furthermore, the investigation of the image states is a powerful tool for probing a variety of physical and chemical phenomena on the nanometer scale. For instance, for metallic vicinals both the confinement and superlattice effects can produce image-band splittings and anticrossings from lateral back scatterings at the step edges as predicted theoretically within a static impenetrable surface model~\cite{lei}. Therefore, to gain insights in the electronic motions and excitations in such nano-materials, theoretical methods are necessary that simulate the details of the processes in the band structure enriched by modifications of the surface. One relatively simpler, and computationally tractable, way to do this is to simulate the motion of electrons transferred from a scattered negative ion as accomplished in our recent publication~\cite{shaw18}. These results may likely be probed experimentally and be utilized to guide future theoretical studies. 

The charge transfer interaction dynamics of an atomic or molecular ion with a surface is highly sensitive to the surface electronic band structure. This fundamental process in ion-surface interaction is valuable to understand the dynamics in processes of scattering, sputtering, adsorption, and molecular dissociation~\cite{rabalais03}. From the applied interests, this mechanism is a crucial middle step in (i) analysis, characterization, and manipulation of surfaces~\cite{stout00}, (ii) semiconductor miniaturization and the production of self-assembled nanodevices~\cite{korkin07}, and (iii) micro-fabrication based on reactive ion etching and ion lithography~\cite{campbell01}. In the recent past, electron transfer between various atomic species with surface containing nanosystems has become a topic of interest, probing effects of the nanosystem's size and shape to determine their electronic structures~\cite{canario03}. 

The energy conserving transfer of an active electron, the resonant charge transfer (RCT), occurs when a near-degeneracy is achieved between a shifted ion affinity-level and various surface localized states. This enables the transfer to (from) an unoccupied (occupied) resonant state of the substrate through wavefunction overlaps. The RCT process in ion-scattering has been the focus of experiments on mono- and polycrystalline metal surfaces~\cite{bahrim, yang, hecht, guillemot, sanchez}. A recent theoretical research studied the RCT interactions of negative ions with nanoisland films~\cite{gainullin15}. Also, a wave packet propagation method was used to access RCT dynamics between excited states of Na nanoislands on Cu(111)~\cite{hakala07}. Treating vicinal steps within a jellium model, H$^-$ neutralization was studied in a wave packet propagation approach, in which steeper-angle scatterings were considered to study the dependence of ion-impact direction vis-a-vis the slope of a local step~\cite{obreshkov06}.  During past years, a full quantum mechanical wave packet propagation approach was employed by our group to conduct detailed RCT studies in ion-scattering and associated neutralization processes in the light of altering crystallographic properties in low Miller index flat surfaces \cite{chak70, chak69, chak241, schmitz}; the results had success in describing some available measurements~\cite{chak69}. For vicinal surfaces with periodic arrays of terraces, the confinement driven reflections of electrons from steps can further modify the free electron dispersion into subband dispersion enriching the RCT process. We recently applied this techniques to study RCT between the H$^-$ ion and vicinally stepped Pd(111) surfaces and computed the ion survival probability \cite{shaw18}. The natural next step therefore is to extend the method to other metallic vicinals to explore general similarities and detailed differences as a function of altering surface band structures. 

In the present study, we accomplish this goal by investigating vicinally stepped Cu(111) and Au(100). The detailed results are presented as a comparative account by including the results of Pd(111) vicinal. Even though we use a rather simple model for the vicinal corrugation, previous utilizations of this model in interpreting angle-resolved photoemission measurements on such surfaces to access surface electronic states~\cite{mugarza,mugarza01} provide confidence in probing, at least qualitatively, the ion-vicinal RCT process in a fully quantum mechanical time propagation treatment. The calculations are done as a function of the component of H$^-$ collision velocity parallel to the surface at shallow incident angles as well as for different distances between the steps on the surface. The electron wave packet probability densities were calculated at all points in space at each time interval. Visualization of these densities indicates that, when the electron transferred from the ion to the metal, it most likely transferred to both the surface and image superlattice states when the ion's approach velocity perpendicular to the surface is slow enough. Two schemes of ion trajectory resulting in different ion-surface interaction times are employed in the simulation to capture the role of the ion ``hangout" time. For a given distance between the steps on the surface, the ion survival probability shows peaks at certain velocities as a robust feature for all three surfaces. It is shown that these peaks appear when the kinetic energy of the electron transferred to an image state matches the subband dispersions supported by the periodic vicinal steps. However, the magnitude of the ion survival is found to sensitively depend on the particular surface band properties, as expected. The result suggest a possibility of accessing superlattice band structures \textit{via} anion-scattering experiments. Atomic units (a.u.) are used in the description of the work, unless mentioned otherwise.
\begin{figure}[h!]
\includegraphics[width=8.3cm]{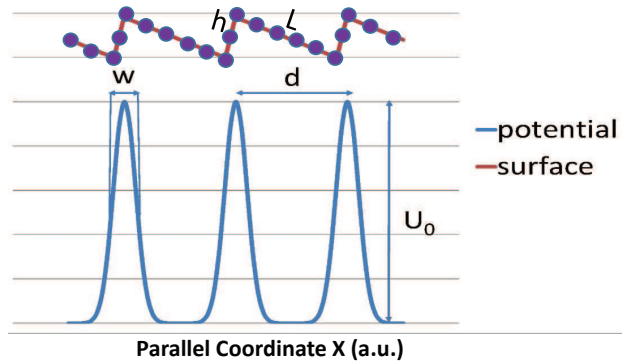}
\centering
\caption{(Color online) The one-dimensional Kronig-Penney potential~\cite{mugarza} and the vicinal surface it models with terrace width ($L$) and step height ($h$). }
\label{fig:fig1}
\end{figure}

\section{Description of the Method}

\subsection{Surface model in vicinal direction}

Due to the repulsive step-step interactions, vicinal corrugations generally appear regularly spaced~\cite{swamy99}. The size of the terrace is determined by the terrace width (vicinal miscut) $L$ and height $h$ of the step. We made use of a two-dimensional model of the metal surface which includes the ion approach direction ($z$) normal to the primordial flat surface and only one direction ($x$) on the flat surface along the vicinal steps. We used the Kronig-Penney (KP) potential to mimic the periodic potential array formed by the step superlattice in the $x$-coordinate shown in Fig.\ 1. The steps in the metal surface are mimicked by the peaks in the KP potential with height $U_0$ and width $w$, and the step separation $d$ (distance between adjacent steps). This model is used as it was successful at describing the experimental photoemission spectra for stepped vicinal metal surfaces \cite{mugarza}. These experiments used vicinal surfaces with the step height of a single atomic layer cut. When fitting the data to the KP model, the largest value of the product $U_0w$ was found to be 0.054 a.u.\ in Ref.\ [\onlinecite{mugarza}]. Therefore, we chose $U_0$ = 0.054 a.u.\ with $w$ = 1 a.u.\ for the results presented in this article. We note that there is no exact correspondence between $U_0$ and $h$. This is because our model does not represent vicinal steps at the atomistic level which needs a full 3D structure calculation. Rather, in the spirit of Ref.\ [\onlinecite{mugarza}], the model mimics the effect of vicinal steps using a flat surface with a potential array for which the strength $U_0w$ correlates the electrostatic strength of a step. We should also note, as discussed in Ref.\ [\onlinecite{mugarza}], that approximating the periodic potential as a Dirac $\delta$-array $\sum _n U_0w \delta(x-nd)$, or other combinations of $U_0$ and $w$ producing the same barrier strength $U_0w$ will not change the model. Indeed, for all our calculations a different combination, namely $U_0$ = 0.027 a.u.\ and $w$ = 2 a.u., fully reproduced the results. And within the same spirit, when the product $U_0w$ was half as great, that is 0.027 a.u., the survival was half as high due to the fact that a weaker barrier increased transmission to the next well reducing the capture rate. We emphasize that in the absence of any parametric form of the vicinal surface potential derived from ab\,initio calculations, we use a scheme of combining this one-dimensional array potential with a known parametric representation based on an ab\,initio flat surface potential as described below.
\begin{figure}[h!]
\includegraphics[width=8.3cm]{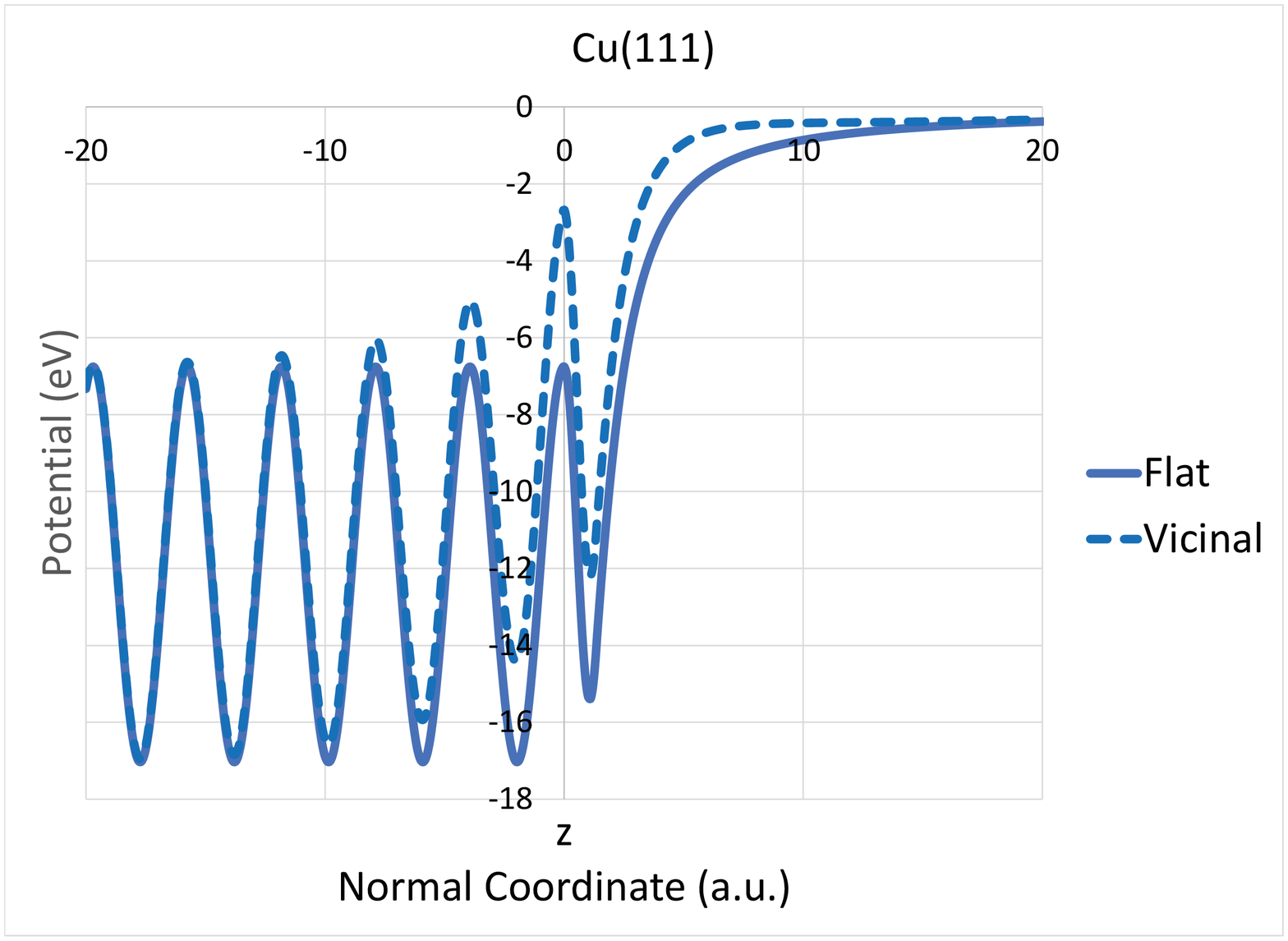}
\centering
\includegraphics[width=8.3cm]{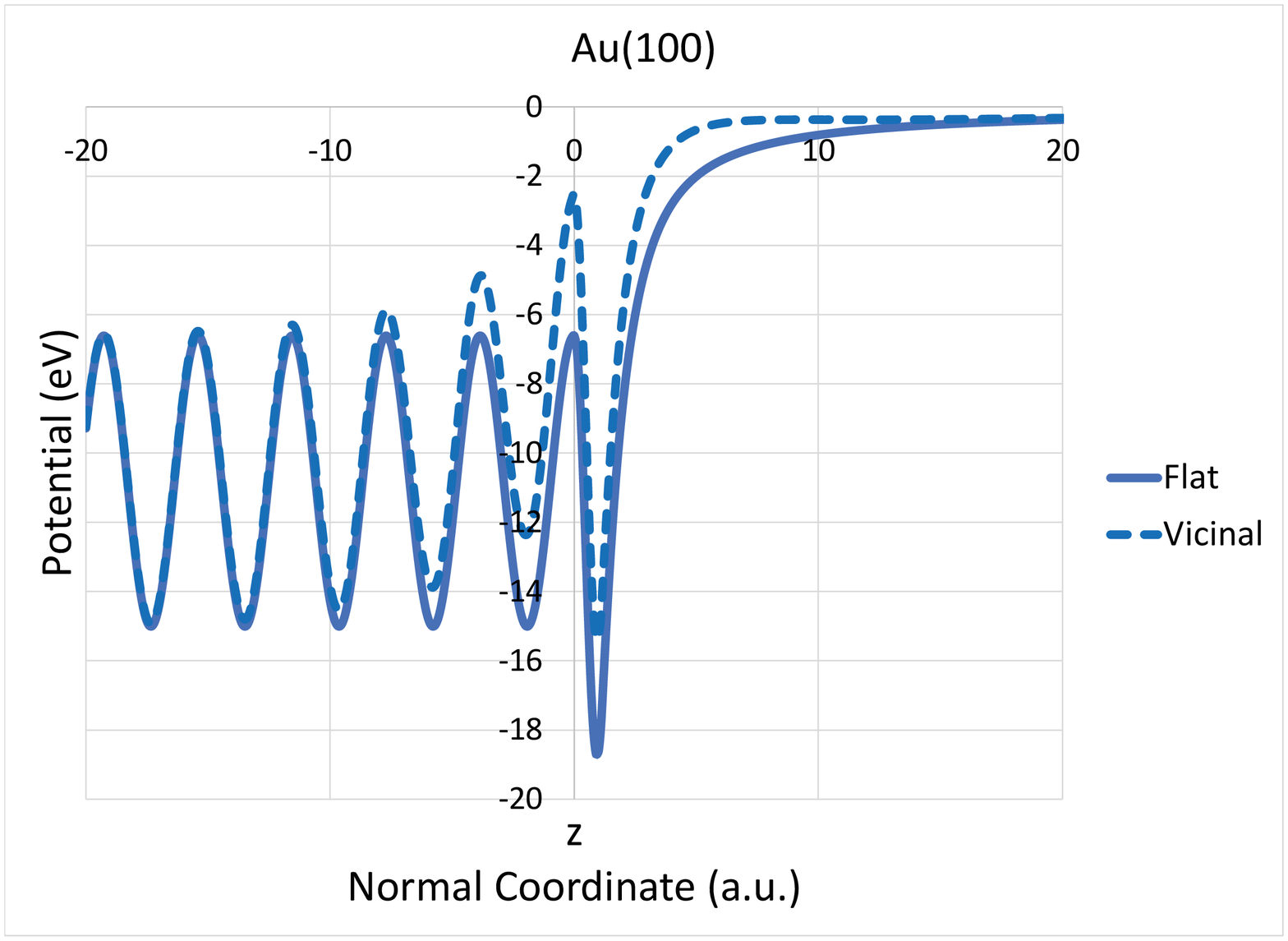}
\centering
\includegraphics[width=8.3cm]{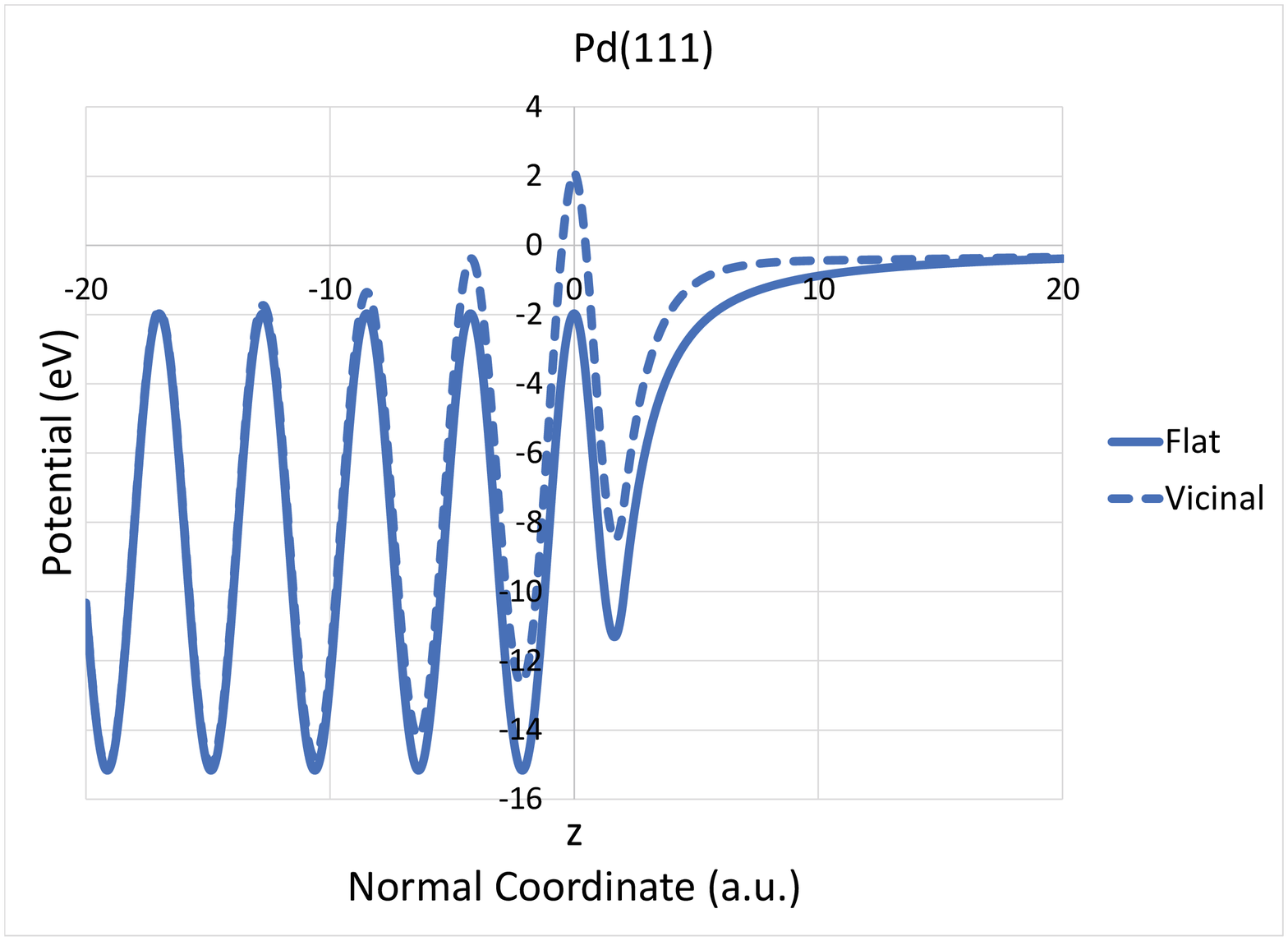}
\caption{(Color online) The potentials in one-dimension for the metals Cu(111) (top), Au(100) (middle) and Pd(111) (bottom). The solid lines are the one-dimensional ($z$) pseudopotentials of the flat surface~\cite{chulkov}, while the dashed lines show the addition of the potential, as in Fig.~1, in the vicinal direction at one of its peaks, scaled by a factor of 2.8 and duly attenuated (see text). The dashed curve for Pd(111) is a $z$-section of Fig.\ 3}
\label{fig:fig2}
\end{figure}
\begin{figure*}
\centering
\includegraphics[width=17cm]{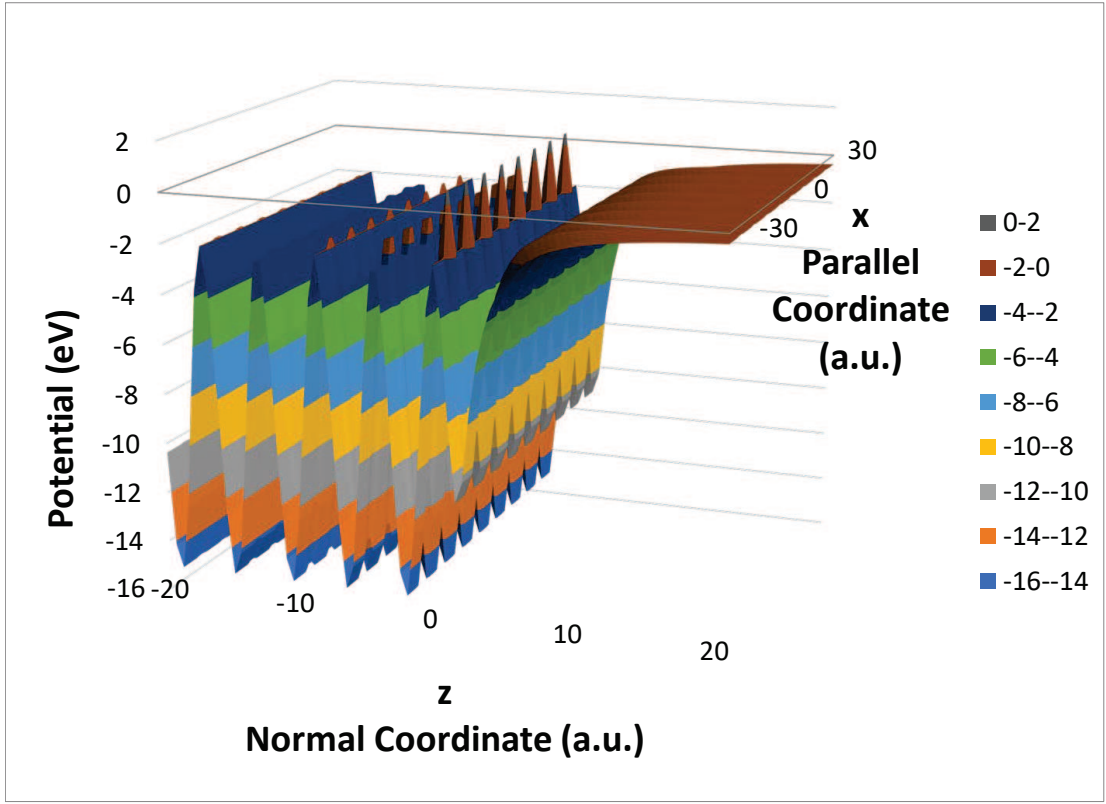}
\caption{(Color online) Schematic of two-dimensional vicinal stepped potential on flat Pd(111) with $d$ being five atomic layers developed using Kroenig-Penny (KP) potential in Fig.\ 1 being super-imposed on one-dimensional pseudopotential of Pd(111)~\cite{chulkov, chulkov97} in Fig.\ 2 (bottom panel). The KP potential is attenuated going far from the precursor flat surface both towards the bulk and the vacuum. To aid visualization, the peaks of the KP potential are scaled by a factor of 2.8 in the figure. Note that the maximum vicinal peaks occur at $z=0$.}
\label{fig:fig3}
\end{figure*}

\subsection{Wave packet propagation}

The details of the propagation methodology are given in Ref.\ \cite{chak70}, and is recently applied to the study of Pd(111) vicinal surface~\cite{shaw18}. The time-dependent electronic wave packet $\Phi\left(\vec{r},t;D\right)$ for the ion-surface combined system is a solution of the time-dependent Schr\"{o}dinger equation
\begin{equation}\label{tdse}
i{\hbar}\frac{\partial}{\partial t}\Phi\left(\vec{r},t;D\right) = H\Phi\left(\vec{r},t;D\right),
\end{equation}
with the general form of the Hamiltonian as
\begin{equation}\label{hamil}
H = -\frac{1}{2}\frac{d^{2}}{dz^{2}}-\frac{1}{2}\frac{d^{2}}{dx^{2}}+V_{\mbox{\scriptsize vi-surf}} (x,z) + V_{\mbox{\scriptsize ion}} (x,z)
\end{equation}
where $D(t)$ is the dynamically changing perpendicular distance between the $z=0$ line (see below) of fixed-in-space metal surface and the ion moving along a trajectory.  The potential, $V_{\mbox{\scriptsize vi-surf}}$, of the vicinal surface will have two components: (i) A one-dimensional potential model in $z$ obtained from pseudopotential local-density-approximation calculations for simple and noble metal surfaces~\cite{chulkov, chulkov97} that represents the flat precursor substrate. This is the model, with free electron motion in the $x$ direction, that was employed in our previous work with flat surfaces \cite{chak70, chak69, chak241, schmitz} and, more recently, with a vicinal Pd(111) surface~\cite{shaw18}. A graph of this potential appears in Fig.\ 2 for each of the three metals. The lattice points of the topmost layer of the substrate is taken at $z=0$ so the peaks in the potential for $z<0$ are the centers of the atomic layers going into the bulk and separated by the interatomic lattice spacing. (ii) Superimposed on (i) is the KP potential model described above that mimics the regular vicinal array of terraces $L$ on the surface along $x$. To limit the vicinal effect far from the surface, the KP potential is attenuated exponentially in both the positive (vacuum side) and negative (bulk side) $z$-direction. Curves representing a section of $V_{\mbox{\scriptsize vi-surf}}$ along $z$-direction through a vicinal peak for each metal are also included in Fig.\ 2. Detailed differences among these curves based on the variation in their flat surface potential properties suggest the role of flat-surface dispersions to also influence corresponding vicinal dispersions. 

Fig.\ 3 provides an illustration of the full 2D $V_{\mbox{\scriptsize vi-surf}}$, only on Pd(111), used in our simulation. The peaks of the KP potential, as seen in both Fig.\ 2 and Fig.\ 3, are set at their maximum (unattenuated) value at $z=0$. This must be the case if the atomic layer positions of the precursor flat surface are not to be altered by the vicinal potential. Therefore, the top atomic layer used in the flat metal potentials in Fig.\ 2 defines $z=0$. Smaller peaks are barely visible in Fig.\ 3 on the top of the potential for the next atomic layer inside the surface which is due to the effect of the exponential attenuation. Small peaks can also be seen at the bottom of the dips. The same attenuation is present for $z>0$ but is not as clearly visible in Fig.\ 3. The 2D potential versions for Cu(111) and Au(100) are qualitatively close to Fig.\ 3, hence not shown.

The pseudopotential shown by the solid line for each metal in Fig.\ 2 requires five parameters to define it~\cite{chulkov, chulkov97}. One of these is the lattice spacing ($a_s$) between atomic planes in the metal. The other four are related to the top and bottom energies of the projected band gap, the surface state energy and the first image state energy~\cite{chulkov97}. Therefore, differences of H$^-$ RCT dynamics between the flat surfaces must be explained by differences between these five quantities for the different metals. Similarly, the surface potential in the vicinal direction is defined by the three parameters shown in Fig.\ 1. Results for a particular vicinally stepped surface, therefore, should modify based on these three parameters as was shown in Ref.\ [\onlinecite{shaw18}]. However, the comparison of vicinal RCT among various metals, the focus of the present study, will partly borrow from their flat-substrate properties as will be shown.

The H$^-$ ion is described by a single-electron model potential, $V_{\mbox{\scriptsize ion}}$, which includes the interaction of a polarizable hydrogen core with the active electron \cite{ermoshin}. However, we employ an appropriately re-parametrized version of this potential~\cite{chak70}, as was also applied in our other publications~\cite{chak69, chak241, schmitz, shaw18}. This form is commensurate with our two-dimensional propagation scheme and produces the correct ion affinity level energy \Eion = 0.0275 a.u. (0.75 eV).

The propagation by one time step $\Delta t$ will yield
\begin{equation}\label{propa}
\Phi(\vec{r},t+\Delta t;D) = exp[iH(D)\Delta t]\Phi(\vec{r},t;D)
\end{equation}
where the asymptotic initial packet $\Phi_{\mbox{\scriptsize ion}}(\vec{r},t=0,D=\infty)$ is the unperturbed H$^-$ wave function $\Phi_{\mbox{\scriptsize ion}}(\vec{r},D)$. The ion-survival amplitude, or autocorrelation, is then calculated by the overlap
\begin{equation}\label{auto}
A(t) = \left\langle \Phi(\vec{r},t)|\Phi_{\mbox{\scriptsize ion}}(\vec{r})\right\rangle.
\end{equation}
We employ the split-operator Crank-Nicholson propagation method in conjunction with the unitary and unconditionally stable Cayley scheme to evaluate $\Phi(\vec{r},t;D)$ in successive time steps \cite{chak70,press}. Obviously, the propagation limits the motion of the active electron to the scattering plane of the ion.

\subsection{Ion trajectories}

We assume that when the ion reflects from the surface, the angle of reflection is the same as the angle of incidence measured from the flat substrate plane. The Hydrogen ion impinges on the surface at shallow angles with respect to $x$. Inputs to the calculation are the component of ion velocity normal ($z$) to the surface, $\vnor$, and the component of velocity parallel ($x$) to the surface, $\vpar$. The computer program aims the ion at a point halfway between two steps as well as directly on a step. For one scheme of trajectory, the incident ion decelerates along the $z$ direction, close to the surface, due to the net repulsive interaction between the ion core and the surface atoms while it stays constant in $\vpar$. For a given initial velocity, we simulate a classical trajectory based on Biersack-Ziegler (BZ) interatomic potentials to model this repulsion \cite{chak70, biersack}. The slowdown of $\vnor$ to zero at the point of closest approach (the turning point) and its subsequent gradual regaining of the initial speed at the initial distance while reflecting symmetrically at constant $\vpar$ makes the resulting trajectory somewhat parabolic. 

In order to identify features of the results that depend on the trajectory, we also employed a basic trajectory that we call a \textit{broken straight line} (BSL) trajectory. In this trajectory the ion approaches along a straight line with constant velocity (zero repulsion) and reflects back along a straight line with the same constant velocity at the same angle to the surface. For BSL, the same distance ($D_{cl}$) of closest approach is used as in the BZ trajectory. Obviously, in the absence of slowing down, the ion on a BSL trajectory will have shorter time of interaction with the surface than on a BZ trajectory. 

The ionic motion on each trajectory is then incorporated in the propagation by adding the translational phase $(\vnor z + \vpar x + v^2t/2)$~\cite{chak70} as well as by shifting the center of the ion potential in \eq{hamil} to follow the trajectory that corresponds to evolving $D(t)$. We note that to formulate the trajectory we calculated the BZ potential as if the surface was flat at $z =0$. But this limitation of our trajectories, blind to vicinal shapes, is not expected to qualitatively affect the main results which should predominantly depend on the RCT process. 

In the grazing scattering, ion neutralization on metal surfaces can have consequence from the so called parallel velocity effect which causes a shift of the Fermi sphere~\cite{borisov96} over a range of $\vpar$.  This effect is significant on cation neutralization that has a strong capture rate from the metal's Fermi sea, while the current process addresses the neutralization of anions. As estimated in Ref.\ [\onlinecite{shaw18}], the Pd Fermi energy ($E_f$) from the bottom of the valence band is about 0.262 a.u.~\cite{mueller70} which corresponds to the magnitude of Fermi wave vector of about $k_f=$ 0.72 a.u., or $k_f=$ 0.85 a.u. after accounting in a 41\% raise in the electron effective mass for Pd~\cite{mueller70}. The situation for Cu ($E_f$ = 0.257 a.u.) and Au ($E_f$ = 0.203 a.u.) will be largely similar. This may imply that the observed Fermi energy $(k_f-\vpar)^2/2$ from the ion's moving frame~\cite{winter91} may not influence the energy range of the current RCT process (which is very close to the metals' image state energies as discussed in the following section) at least up to about $\vpar = 0.5$ a.u.\ within which the strongest structures in the ion survival (Figs.\ 4, 5, and 6) occur. 

In our simulations, $D_{cl}$ to the substrate will be kept fixed. This is reasonable as the initial value of $\vnor$ at $z=20$ a.u.\ will be constant at a small value 0.03 a.u.\ and thereby largely omits the variation of the dynamics as a function of $\vnor$. We are interested in changes of survival probability due to the surface vicinal structure by varying $\vpar$. Long after the ion's reflection, the final ion-survival probability will be obtained by
\begin{equation}\label{surv}
P=\lim_{t\rightarrow\infty}\left|A\left(t\right)\right|^{2},
\end{equation}
which corresponds to fractions of the survived incoming ions that an experiment can measure~\cite{guillemot}.

\subsection{Simulations}

To calculate the final ion-survival probability, the computer program calculates the electron wave packet density at all points in space at each time interval. This data was used to produce detailed animations of the changing electron wave packet density with time. Though the initial and final value of $z$ was fixed at 20 a.u., the initial and final values of $x$ varied with $\vpar$. Due to a small $\vnor$ value and many values of relatively fast $\vpar$, the ion's flyby was nearly grazing the surface. Consequently, the size of the surface, $|x_{\mbox{\scriptsize final}} - x_{\mbox{\scriptsize initial}}|$, entered in to the propagation was large, resulting in the execution time of the computer program to be very long.  As a result, thread-based parallel computing using OpenMP was employed. To accumulate the amount of data we used, parameter sweeps were done with the program on the Stampede supercomputer at the University of Texas at Austin. More than 500 survival probabilities were calculated in this way for each of the stepped metal surfaces considered in this paper. It should be noted that calculations were done in the 2D model described above and therefore all figures present 2D model results.
\begin{figure}[h!]
\includegraphics[width=8.3cm]{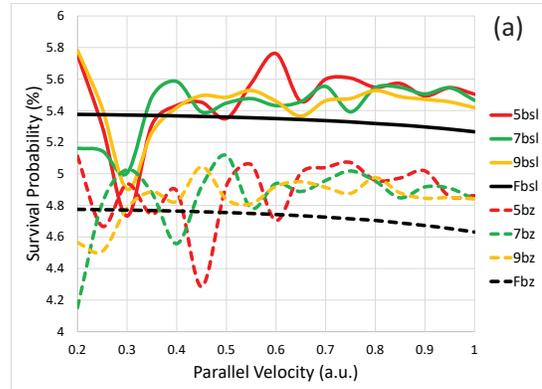}
\centering
\includegraphics[width=8.3cm]{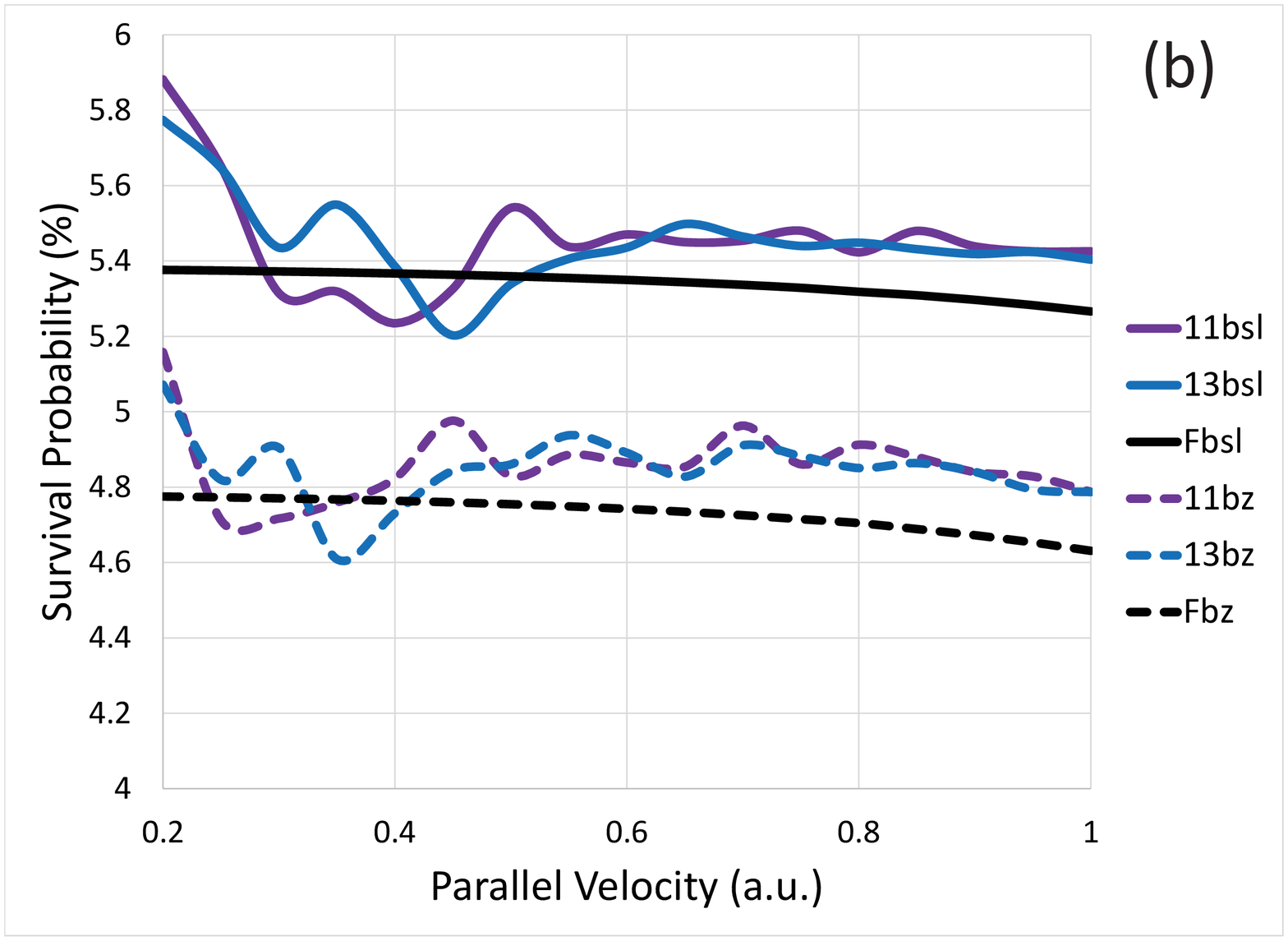}
\caption{(Color online) H$^-$ survival probability for one atomic-layer high vicinally stepped Cu(111) as a function of ion parallel velocity ($\vpar$) as H$^-$ approaches the center of a terrace for the BZ trajectory (solid line) and the BSL trajectory (dashed line) for inter-step distance ($d$) of 5, 7, and 9 lattice spacings (a) and for $d$ of 11 and 13 lattice spacings (b).}
\label{fig:fig4}
\end{figure}
\begin{figure}[h!]
\includegraphics[width=8.3cm]{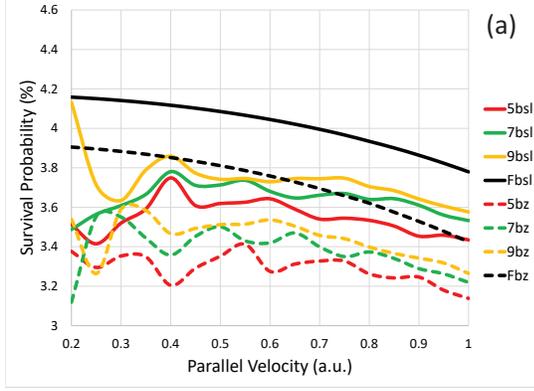}
\centering
\includegraphics[width=8.3cm]{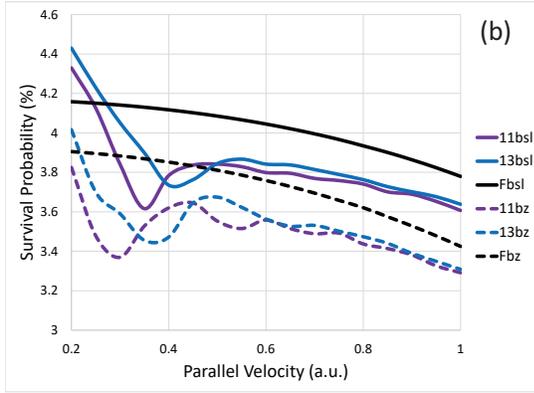}
\caption{(Color online) Same as Fig.\ 4, but for vicinally stepped Au(100).}
\label{fig:fig5}
\end{figure}
\begin{figure}[h!]
\includegraphics[width=8.3cm]{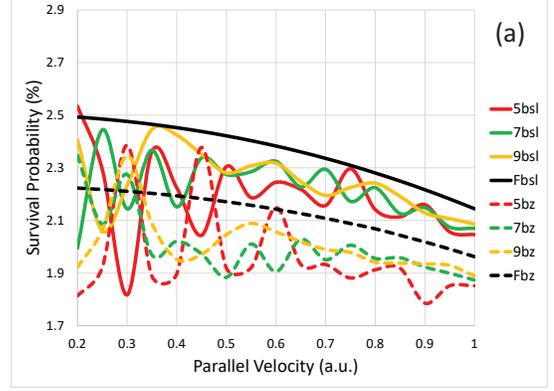}
\centering
\includegraphics[width=8.3cm]{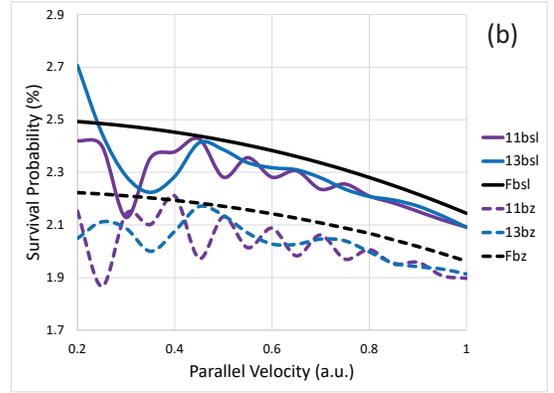}
\caption{(Color online) Same as Fig.\ 4, but for vicinally stepped Pd(111).}
\label{fig:fig6}
\end{figure}

\section{Results and Discussions}

\subsection{Ion survival}

H$^-$ survival probabilities on vicinally stepped Cu(111), Au(100), and Pd(111), after the ion returns to the initial $z$ = 20 a.u.\ $(t\rightarrow\infty)$, were calculated for parallel velocities, $\vpar$, ranging from 0.2 to 1.0 a.u.\ in steps of 0.05 a.u. This is equivalent to a range of ion scattering angles from 8.53 to 1.72 degrees with respect to the surface. The survival probabilities were calculated for the distance between steps ($d$) of 5, 7, 9, 11 and 13 lattice spacings ($a_s$). These lattice spacings were $a_s$ = 3.94 a.u.\ for Cu(111), $a_s$ = 3.853 a.u.\ for Au(100) and $a_s$ = 4.25 a.u.\ for Pd(111). The results of the hydrogen ion survival probability are shown in Figs.\ 4, 5, and 6 for two different trajectories: BZ (solid line) and BSL (dashed line). The graphs were fit to the calculated data points using a cubic spline. The ion survival result, labeled ``F'' in the legend, in front of the flat surface under the same kinematic conditions is also shown for comparison. As seen, the vicinal steps cause significant modulations in the ion survival probability, whereas the flat surface result is smooth. 

In principle, the ion survival on a flat surface should be smooth as a function of $\vpar$ for a fixed $\vnor$ owing to the parabolic free electron dispersion of a flat surface. Moreover, for a flat surface with a projected band gap in the direction normal to the surface, like in Pd(111)~\cite{chulkov}, that resists decay in the $z$-direction, the survival should be almost steady as a function of $\vpar$. However, our flat surface results in Figs.\ 4, 5 and 6 are showing a slow monotonic decrease which is likely because of a numerical artifact due to slightly imperfect boundary absorbers used in the simulation to mimic an infinite surface. It may also partly be due to slow ``evaporation" of the electron probability along the image state Rydberg series to the vacuum. It is then expected that this overall background loss also exists for various vicinal surface results, but that does not alter the main results of modulations in the ion survival arising from the RCT electrons' access to the subband structure. The reason for this is that the error introduced by the boundary absorbers is a systematic error. It will effect all results by the same degree. This will not change the position of the peaks in the survival probability, only their relative amplitudes. It can be seen in Figs.\ 4, 5 and 6 that the flat surface survival probability is generally greater than the vicinal surface survival probability for Au(100) and Pd(111) but smaller for Cu(111) -- a trend true for both the BZ and BSL trajectories. This may be due to the fact that the textured surfaces are modifying the energies of the band gap, the surface state and the image states represented by the parameters of the pseudopotential we are using to model. 

It was found in our previous publication~\cite{shaw18} that the distance $d$ between vicinal steps is one important surface structure property that effects the position of the modulation peaks. As mentioned earlier, $D_{cl}$ = 1 a.u.\ in all results presented here. For larger $D_{cl}$ to a given surface (results not shown) the variations in survival probability became smaller which is simply due to the fact that a distant ion feels the vicinal steps weakly. The location of the peaks also slightly changed with different $D_{cl}$ which is connected to the altering ion-surface effective interaction time - the time of being close enough to the surface. The shorter the $D_{cl}$, the longer is the interaction time. A longer interaction time will allow the ion affinity level to adiabatically shift more in energy~\cite{chak70} causing the electron to land in a KP potential well with a slightly different speed than that for a shorter interaction time. This can offset a little the peak positions between different $D_{cl}$ which will be clear from the discussion in subsection IIID. The interaction time effect also governs the differences between the results from BZ versus BSL trajectories in Figs.\ 4, 5 and 6. The same $D_{cl}$ = 1 a.u.\ for both trajectories enables the ion to sense the steps almost equally resulting, roughly, in the similar strength of modulations irrespective of the trajectory. Note however, in all the results, a slowing ion on the BZ trajectory with a longer interaction time produces consistently lower survival rates than that of an ion on a BSL track, since longer times facilitate higher decays. On the other hand on a BSL path, the affinity level has slightly less time to shift causing small mismatches in peak positions between BSL versus BZ; we will return to this point again in subsection IIID.

There is more. The lattice spacing $a_s$ of 3.94 for Cu(111) and 3.853 for Au(100) are very close indicating that the true lengths of $d$ are effectively same for these surfaces. In spite of this, as can be noted between Figs.\ (4) and (5), details of the ion survival, even for a given ion trajectory, are very different for these two metals suggesting that the structures are dependent on metal type as well, and not just on $d$.  Approximating the wave packet $\Phi(\vec{r},t)$ on a vicinal surface to consist of a flat substrate and a vicinal step component, we can write the autocorrelation as,
\begin{eqnarray}\label{auto-comp}
A(t) &=& \left\langle \Phi_{\mbox{\scriptsize flat}}(\vec{r},t) + \Phi_{\mbox{\scriptsize step}}(\vec{r},t)|\Phi_{\mbox{\scriptsize ion}}(\vec{r})\right\rangle \nonumber \\
     &=& A_{\mbox{\scriptsize flat}}(t) + A_{\mbox{\scriptsize step}}(t).
\end{eqnarray}
Upon inserting \eq{auto-comp} in the definition of survival probability \eq{surv} we can write
\begin{equation}\label{surv-comp}
P=\lim_{t\rightarrow\infty}\left[\left|A_{\mbox{\scriptsize flat}}\right|^{2} + \left|A_{\mbox{\scriptsize step}}\right|^{2} + A_{\mbox{\scriptsize flat}}A_{\mbox{\scriptsize step}}^{\ast}+A_{\mbox{\scriptsize flat}}^{\ast}A_{\mbox{\scriptsize step}}\right]
\end{equation}
to approximate the ion survival on a vicinal surface. Obviously, the first term on the right side of \eq{surv-comp} suggests that the non-modulating average of the survival is a direct substrate property. The leading contribution to the modulations derives from the second term that determines the peak positions from confinements between vicinal steps. It will be shown in subsection IIID that these positions depend on image state energies which are slightly different from one metal to another. However, the last two terms together, embodying a pure interference between the substrate and steps, is responsible for the detailed shape and magnitude differences in survival structures for vicinals among various metals. Evidently, the dispersion energetics of precursor flat surface is important.
\begin{figure*}
\centering
\includegraphics[width=15cm]{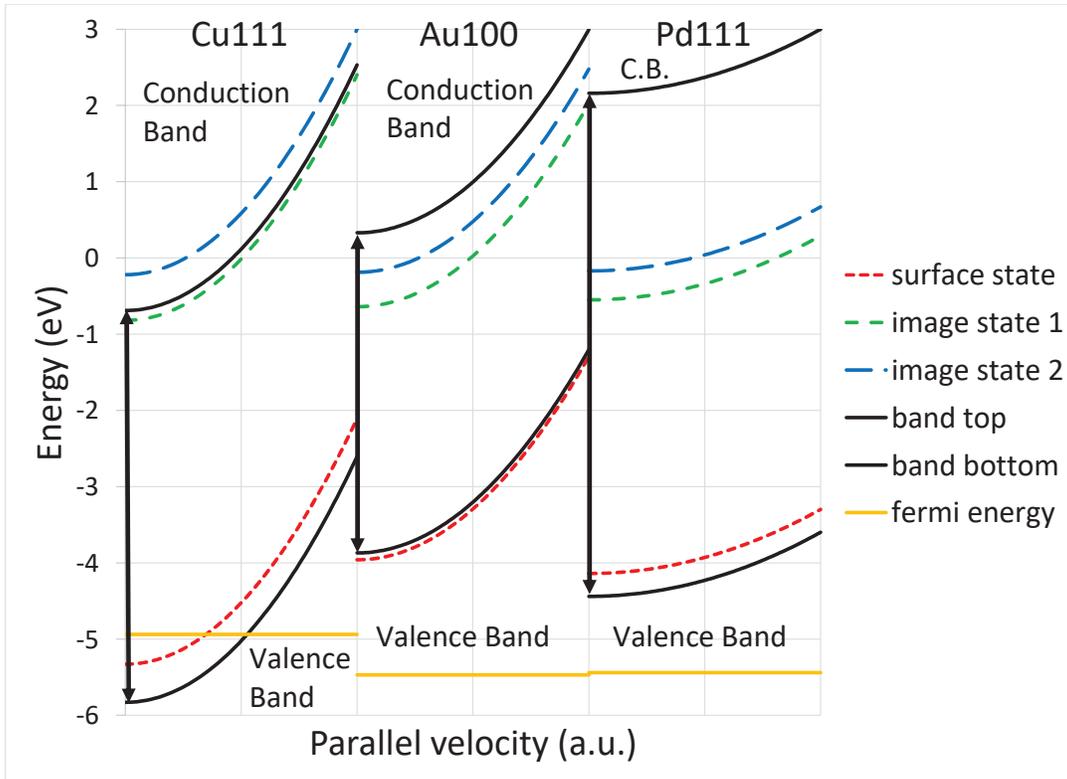}
\caption{(Color online) Dispersion relations for flat Cu(111), Au(100) and Pd(111) substrates assuming a translationally invariant surface and an electron velocity component in the surface plane. We show the Fermi energies only for completeness.}
\label{fig:fig7}
\end{figure*}

\subsection{Effects of dispersion energetics of substrates}

Fig.\ 7 shows the parabolic dispersions for flat substrates featuring the upper and lower edge of the projected band gap, the Shockley surface state and the first and second image states~\cite{chulkov}. It is expected that the incoming ion at its closest approach will resonantly populate both the surface and image state bands~\cite{schmitz, osma, gao}. Comparing Figs.\ 4, 5, and 6 it is seen that the metal with the greatest and smallest survival probabilities are respectively Cu(111) and Pd(111) with Au(100) being between them. This result can be understood from the relative dispersions in Fig.\ 7. As we discussed earlier for Pd(111)~\cite{shaw18}, the electron transferred to the surface state moves away too quickly to be recaptured by the ion, while the ion can primarily recapture from the image states. The question is, what is the probability that the electron in an image state will transfer back to the ion rather than will decay to the metal? This will depend on: (i) Number of available metal states for the electron to transfer to and (ii) the relative transition probability for electron transfer to each of these states versus the ion state. The energy of the electron in the ion is -0.75 eV, which is very close to the energy of the first image state -0.82 eV for Cu(111), -0.64 eV for Au(100) and -0.55 eV for Pd(111). As a result, there is a substantial probability for the electron to transition from an image state back to the ion. It is the large phase space of metal states (collectively the surface state, and valence and conduction band states) which results in survival probabilities of 5\% or less for the ion, as seen in Figs.\ 4, 5, and 6. The size of this phase space for the valence band and conduction band will depend on the location of the band-gap edges. The lower the bottom of the gap, the smaller is the phase space in the valence band. Likewise, the higher the top of the band gap, the smaller is the phase space in the conduction band. Since the valence band states are of lower energy than the first image state, whereas the conduction band is higher in energy than the first image state, the transition probability for the electron in an image state to drop to the lower states, which are the valence band and surface states, are greater than to the conduction band states. The conduction band phase space is further limited by the kinetic energy of the electron, since the conduction electron cannot transition to an energy higher than it's kinetic energy. In general, the larger the energy difference between the image state and the gap bottom, the smaller is the phase space of the valence band and the smaller is the transition probability from the image state to the valence band.

Let us now get more specific. Cu(111) on Fig.\,7, having the largest difference between the first image state energy and the bottom of the band gap, has the highest average H$^-$ ion survival (Fig.\ 4). Comparing Au(100) to Pd(111) in Fig.\ 7, we see that Pd(111) has a slightly larger difference between the first image state and the gap bottom. Furthermore, the overall band gap is larger for Pd(111) than for Au(100). This means that the phase space of metal states is larger for Au(100) than Pd(111) and so one may expect the electron to decay to the bulk more often for Au(100) than Pd(111), since greater the total number of states there are in the metal, the greater the probability that the metal will win the tug-of-war for the electron in the image state~\cite{chulkov, chulkov98}. However, we find Au(100) has the larger average ion survival, comparing Fig.\ 5 with 6! This is due to the fact that the surface energy is in the valence band for Au(100) and in the band gap for Pd(111). If the surface state is in the valence band then it is degenerate with the bulk states of the metal. This can be seen in the images of Fig.\ 8(c). If the surface state is in the band gap, then it adds to the available metal states and, as seen in Figs.\ 8(a), 8(b) and 8(d), is a preferred  energy state. The presence of the surface state in the band gap for Pd(111) gives the metal the edge in its tug-of war-for the electron as compared to Au(100). The surface state is also in the band gap for Cu(111), but this does not overcome the effect of the extremely low bottom of the gap. In any case, as pointed out, this account must be combined with the picture of vicinal confinement to fully understand the results.  
\begin{figure*}
\centering
	\begin{minipage}[b]{8.3cm}
	\includegraphics[width=8cm]{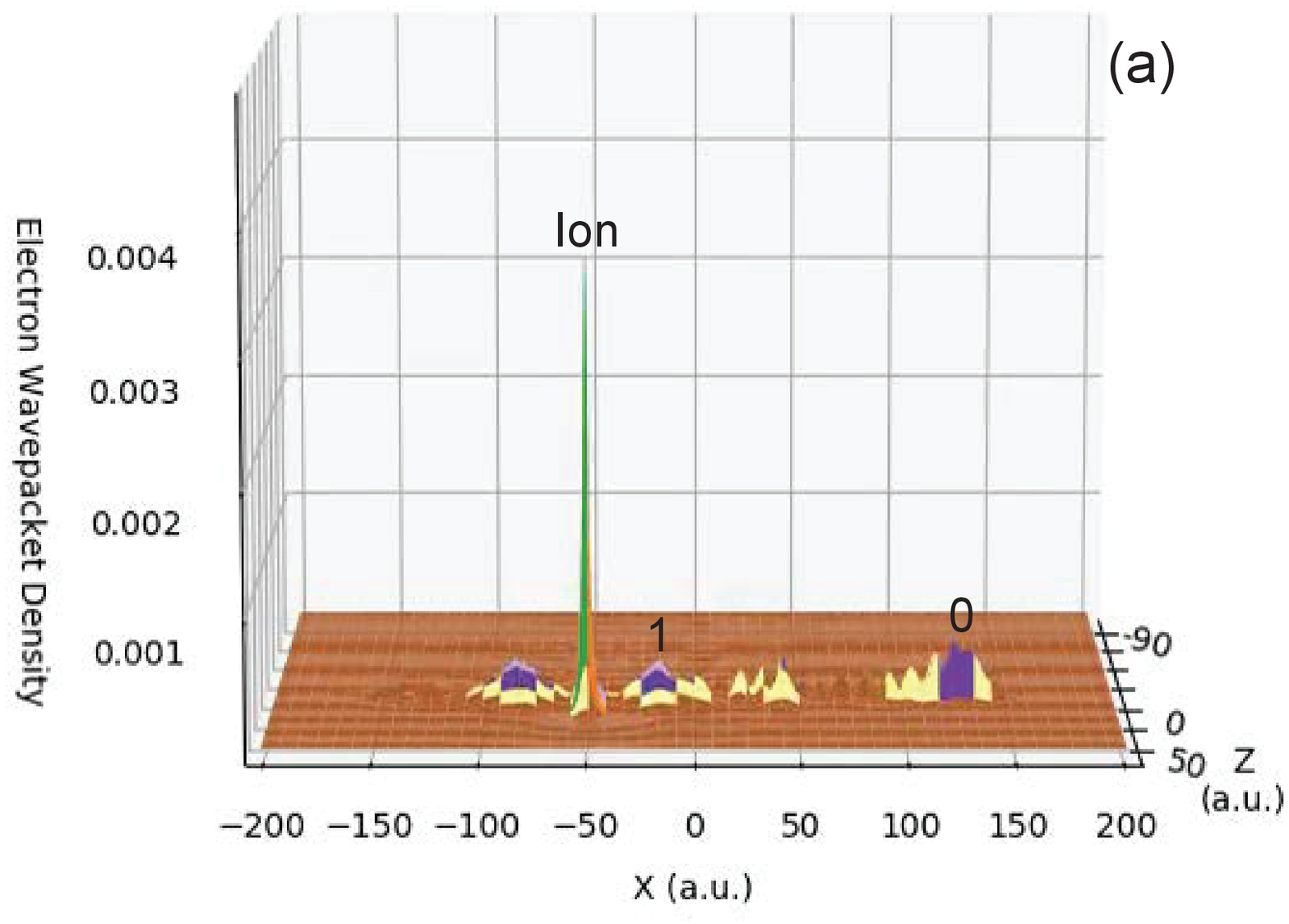}
	\end{minipage}\quad
	\begin{minipage}[b]{8.3cm}
	\includegraphics[width=8cm]{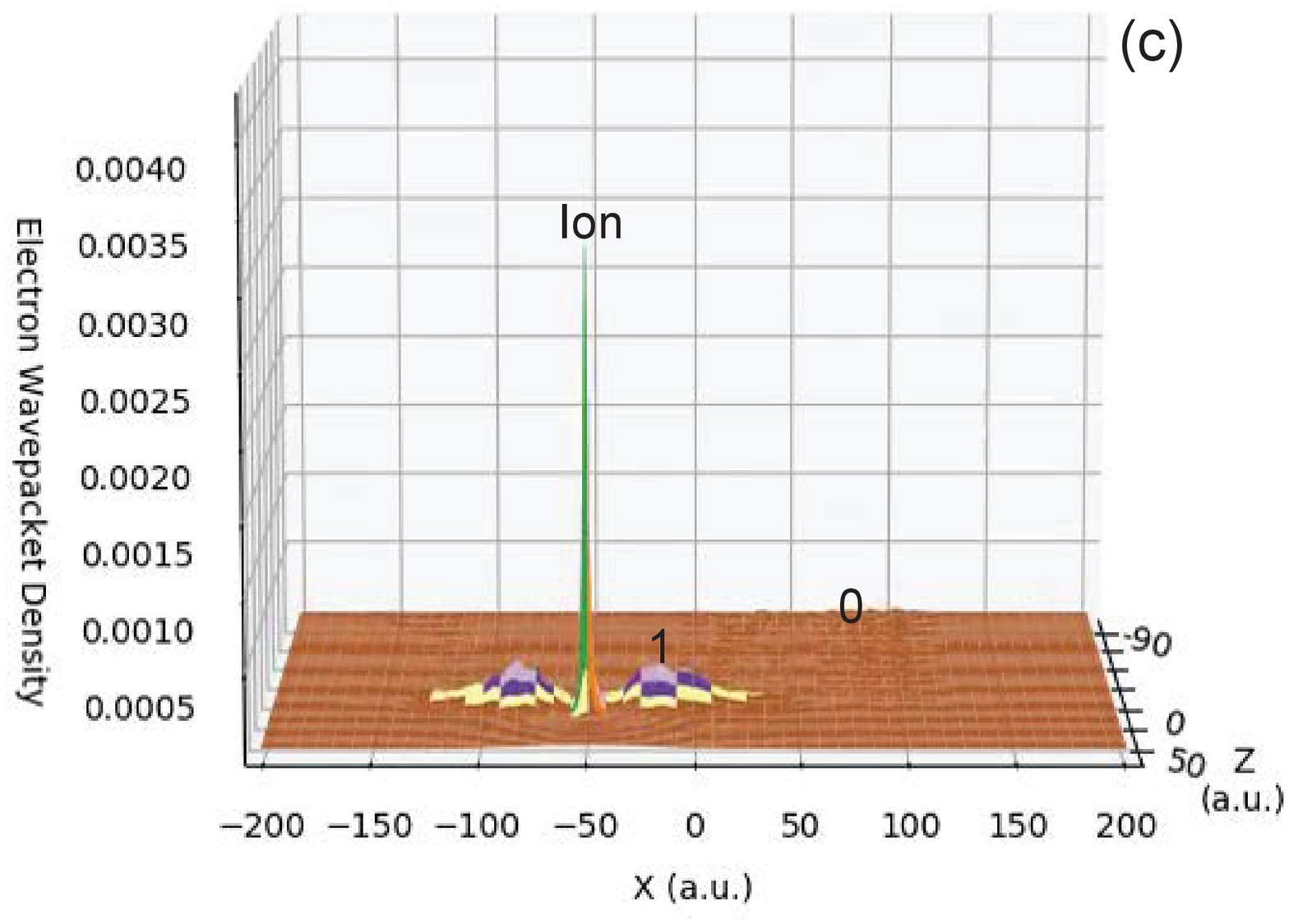}
	\end{minipage}\quad
	\begin{minipage}[b]{8.3cm}
	\includegraphics[width=8cm]{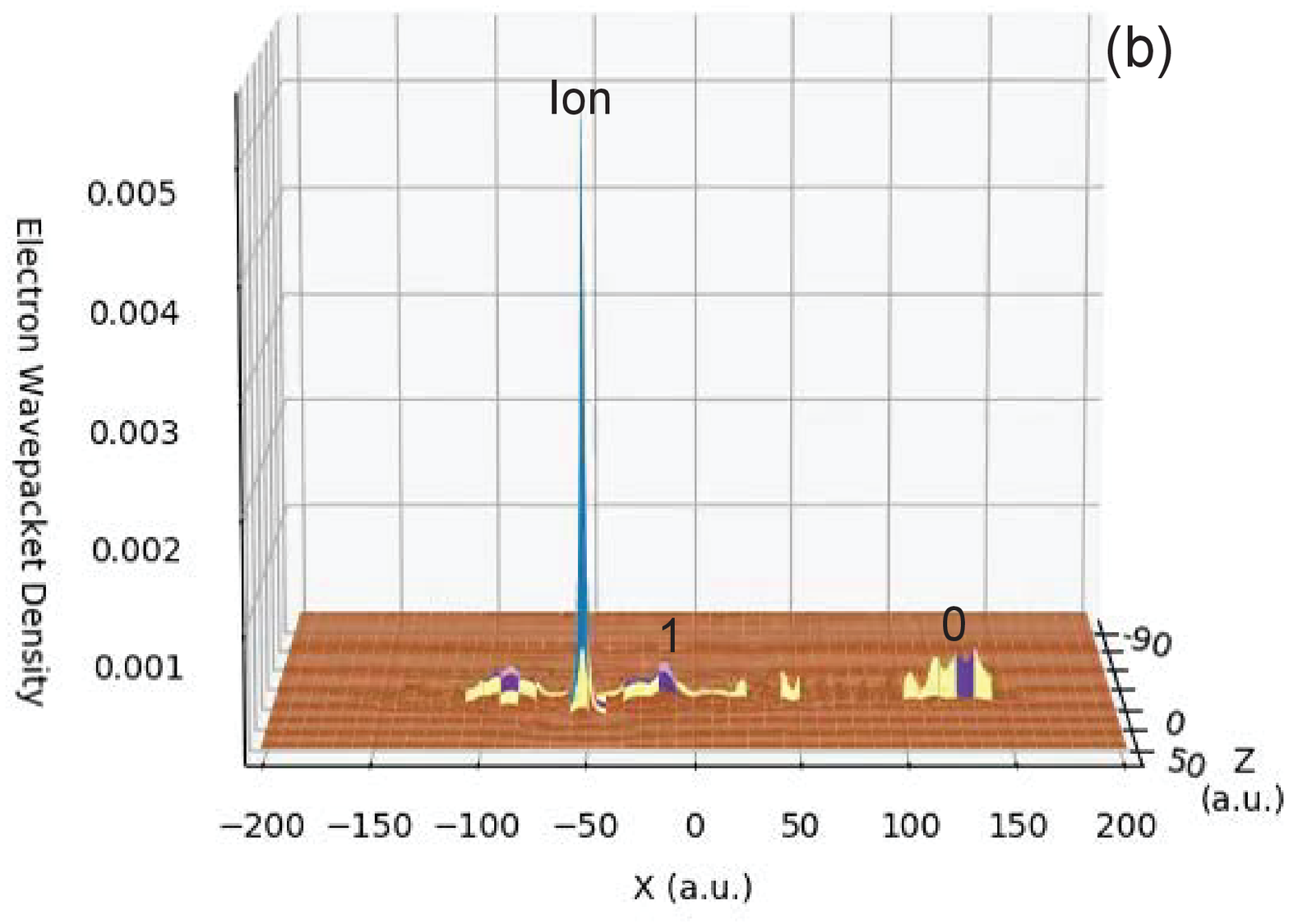}
	\end{minipage}\quad
	\begin{minipage}[b]{8.3cm}
	\includegraphics[width=8cm]{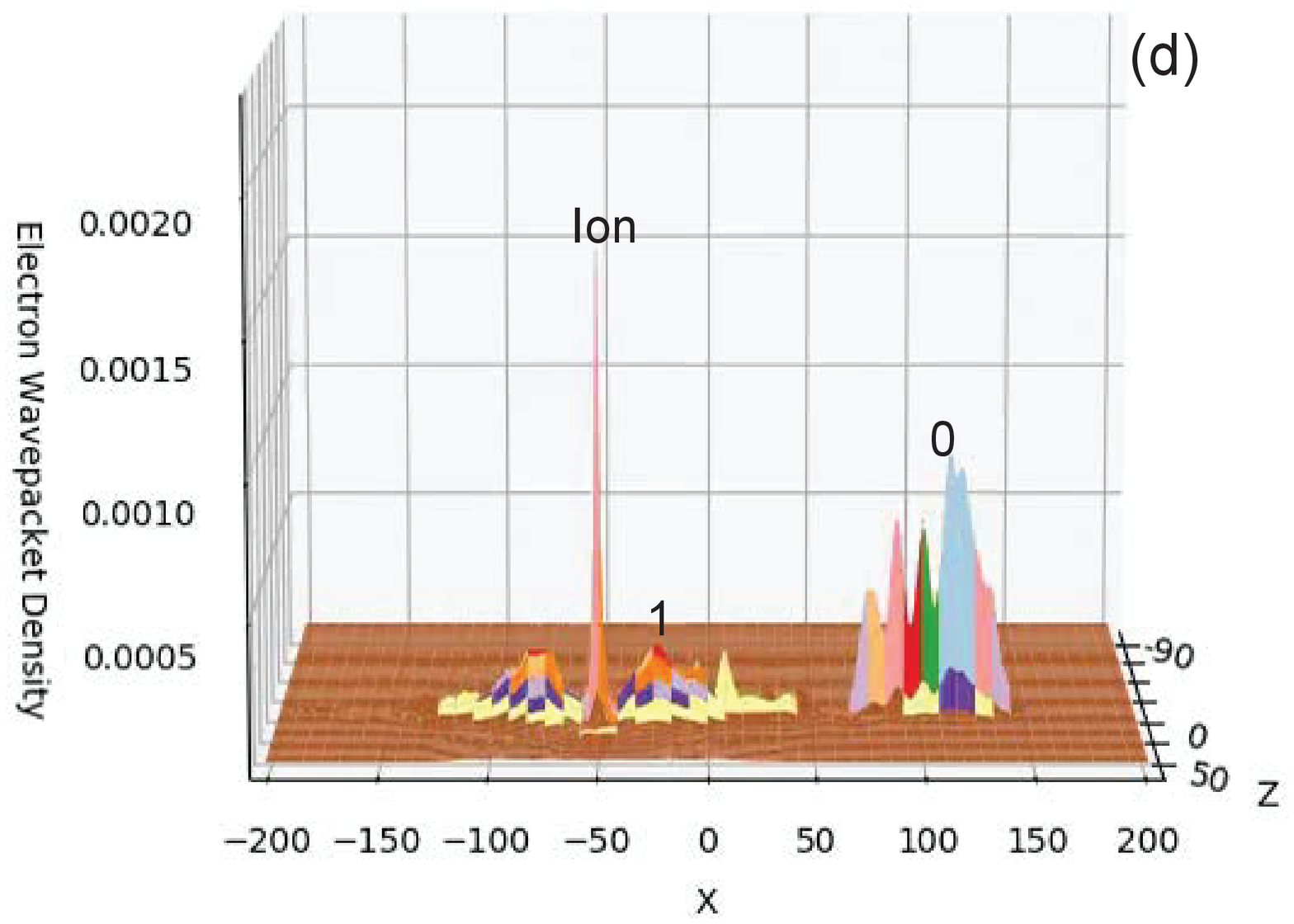}
	\end{minipage}
\caption{(Color online) Time-snapshots of the electron wave packet density: (a) for the vicinal Cu(111) of $d$ =5$a_s$ at the time $t$ where the ion strikes the center of the step; (b) for the vicinal Cu(111) of $d$ =5$a_s$ at the same instant of (a) but where the ion strikes the peak of the step; (c) for the vicinal Au(100) of $d$ =5$a_s$ at the same instant of (a) where the ion strikes the center of the step; (d) for the vicinal Pd(111) of $d$ =5$a_s$ at the same instant of (a) where the ion strikes the center of the step. The electron wave packet density is a dimensionless fraction which is normalized to unity over all space.}
\label{fig:fig8}
\end{figure*}

\subsection{Wave packet density dynamics}

The ion survival amplitude, \eq{auto}, dynamically evolves as the ion approaches the surface. When the ion is close to the surface, there are one or more positions for which the amplitude is found zero. This means that at close distances the electron transfers back and forth between the ion and the metal. Therefore, it is important to probe the RCT dynamics involving the metal states that the ion populates. This was done in detailed in our previous publication~\cite{shaw18}. The vicinal texture is expected to induce superimposed subband modulations to produce survival peaks in Figs.\ 4, 5 and 6. These survival peaks and valleys are just what one would expect from an interference pattern produced by probability waves reflecting back and forth between the vicinal steps of the surface. These interference patterns can be seen in the surface and image state pulses (labeled 0 and 1 respectively) shown in the images of Fig.\ 8 for the different metals. Fig.\ 8 shows snapshots, all at the same instant and for $d$ =5$a_s$, from animations produced of the wave packet probability density as a function of time. For this figure, the ion starts at the right side of the graph and heads into the paper toward a surface and to the left, approaching closest to the surface at $x$ = 0. The surface of the metal is at $z$ = 0. For Cu(111), an image, Fig.\ 8(b), is included for when the ion is aimed at the step whereas for the other images the ion is aimed at a point midway between the steps. Note that for Au(100), Fig.\ 8(c), the surface state peaks are decaying into the bulk of the metal whereas the surface state peaks for Cu(111) and Pd(111), Figs.\ 8(a), 8(b) and 8(d), remain at the surface of the metal. This can be understood from the dispersions shown in Fig.\ 7. For Au(100) the surface state energy is in the valence band and so the electron decays from the surface state into the bulk of the metal. For Cu(111) and Pd(111) the surface state energy is in the band gap so the electron stays in this surface state much longer. The image state stays just outside the surface for all three metals as the image state energies are all within the band gap with the exception of the second image state for Cu(111) which is in the conduction band. Therefore, an electron captured in the second image state would quickly decay into the bulk of the metal via the conduction band. 
\begin{figure*}
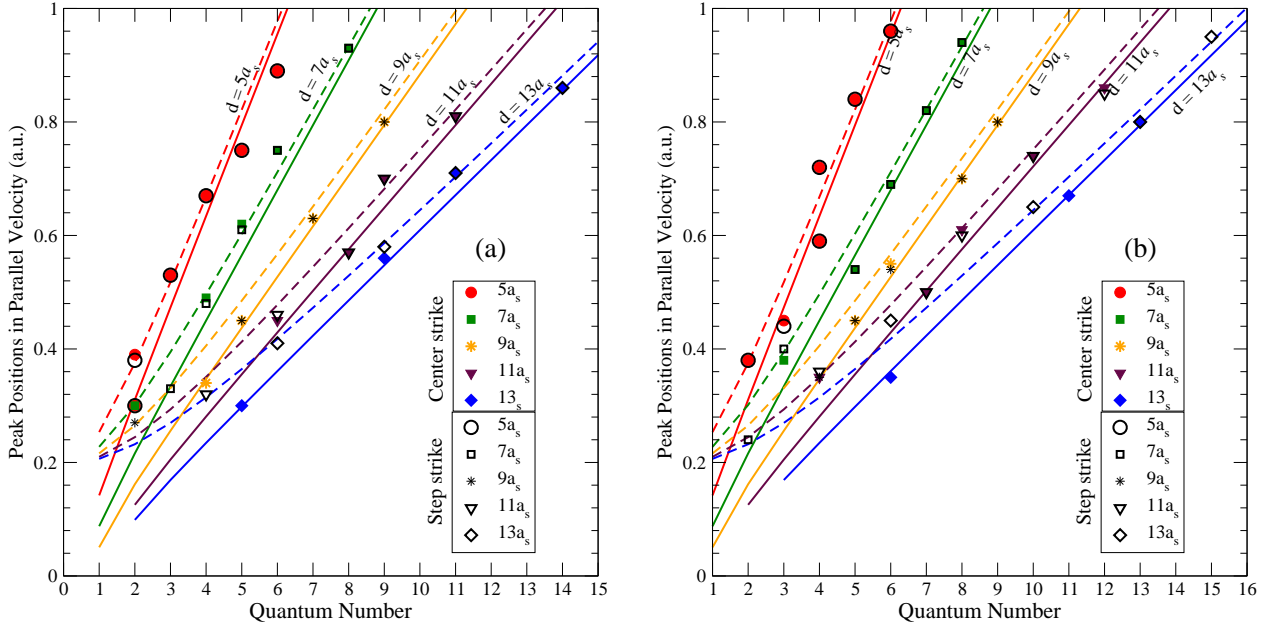

\centering
	\begin{minipage}[b]{8.3cm}
	\includegraphics[width=8cm]{Cu111BZ.eps}
	\end{minipage}\quad
	\begin{minipage}[b]{8.3cm}
	\includegraphics[width=8cm]{Cu111BSL.eps}
		\end{minipage}
\caption{(Color online) Comparison of the ion survival peak positions in the parallel velocity scale for various values of vicinal step separation $d$ on Cu(111) with an analytic square well potential model, using the energy of the first (solid line) and second (dashed line) image state, and  for the BZ trajectory (left) and the BSL trajectory (right).}
\label{fig:fig9}
\end{figure*}
\begin{figure*}
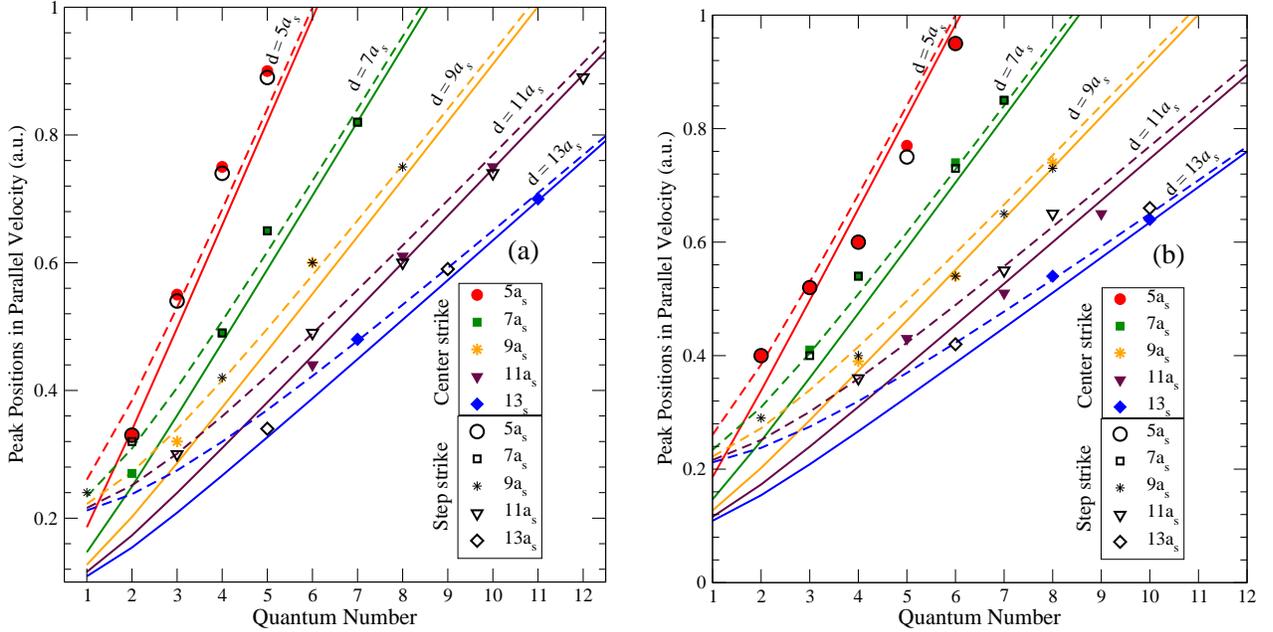

\centering
	\begin{minipage}[b]{8.3cm}
	\includegraphics[width=8cm]{Au100BZ.eps}
	\end{minipage}\quad
	\begin{minipage}[b]{8.3cm}
	\includegraphics[width=8cm]{Au100BSL.eps}
		\end{minipage}
\caption{(Color online) Same as Fig.\ 9, but for Au(100) vicinal surface.}
\label{fig:fig10}
\end{figure*}
\begin{figure*}
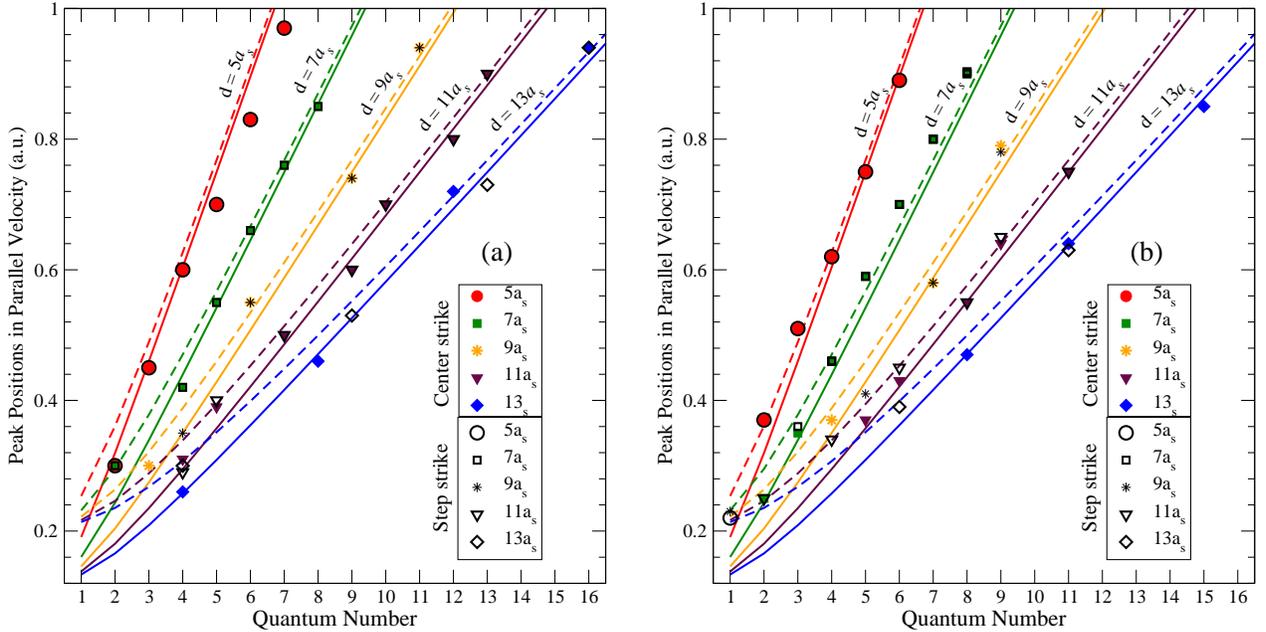

\centering
	\begin{minipage}[b]{8.3cm}
	\includegraphics[width=8cm]{Pd111BZ.eps}
	\end{minipage}\quad
	\begin{minipage}[b]{8.3cm}
	\includegraphics[width=8cm]{Pd111BSL.eps}
		\end{minipage}
\caption{(Color online) Same as Fig.\ 9, but for Pd(111) vicinal surface.}
\label{fig:fig11}
\end{figure*}

\subsection{Superlattice states from lateral confinement}

As discussed in Ref.\ [\onlinecite{shaw18}], the periodic array of potential barriers with finite $U_ow$ and period step separation $d$, representing the vicinal terraces, will induce both reflections and transmissions, respectively from and through the steps, of electrons propagating along the surface. This will give rise to subbands in $x$-direction with the reciprocal vector $2\pi/d$, which zone-folds~\cite{mugarza,shaw18} the surface and image states subbands, which are populate. We demonstrated earlier~\cite{shaw18} and also noted in the current discussion that the depleted ion has the greatest likelihood to recapture electron probability from the lowest image state bands. It is therefore expected that roughly whenever the ion's parallel energy will intersect a subband image state dispersion a resonance-type condition is reached, and the recapture rate by the ion will increase. This will produce peaks in the survival probability, as our main results in Figs.\ 4, 5, and 6 show. 

One simple way to verify that these subband states from surface-parallel confinements induce survival peaks is to compare our numerical results with the analytic infinite-barrier square-well model. This is because ultimately in the infinite barrier limit, these subband dispersions will turn into flat quantum levels as shown in our previous publication~\cite{shaw18}. Therefore, this can still qualitatively guide us in capturing the positions of the survival peaks as we shall show below. For the infinite square well potential, the allowed electron de Broglie wavelengths are given by $\lambda_{n}=2d/n$, where $n$ is a positive non-zero integer we shall call the quantum number and $d$ is the width of the well. From this, it is straightforward to show that the quantized kinetic energy of an electron in the well, in atomic units, is $\left(n \pi/d \sqrt{2}\right)^{2}$. By RCT energy conservation, setting this kinetic energy to be equal to the ion's parallel kinetic energy $(\vpar)^2/2$, {\em plus} the transition energy from the ion level to an image level, one solves for $\vpar$ as 
\begin{equation}\label{sq-well}
\vpar=\sqrt{2\left((n \pi/d \sqrt{2})^{2}+\Eion-E\right)}
\end{equation}
that will give standing electron probability density, where $E$ is an image state energy for the flat surface. All quantities are in atomic units. In \eq{sq-well}, we entirely neglected the ion kinetic energy in the normal direction since the $\vnor$ value chosen is miniscule, and also disregarded the shift of the ion level. 

\eq{sq-well} is plotted for the first (solid lines) and second (dashed lines) image state energies in Figs.\ 9, 10 and 11 for five values of $d$ considered in this work. The values of $\vpar$ for which peaks occur in Figs.\ 4, 5 and 6 are also plotted as symbols on these graphs for simulations when the ion aims both at the center of a terrace (solid symbols) and the top of a step (hollow symbols). The vast majority of the symbols fall reasonably close to corresponding lines obtained from the model, guiding the broad underlying physics. There are a few symbols which are less close to a line than others. The level of agreement is, however, remarkable, given the simplicity of the model. This is particularly true due to the realities in our simulation that (i) the ion actually interacts with a subband dispersion process but not with an ideal well of infinite height and (ii) the image population density spreads over the entire image state Rydberg series inducing some uncertainty in the transition energy. 

Some detailed observations can be made in Figs.\ 9, 10 and 11. The pattern of the quantum numbers of the peaks is what we expect. When $d$ increases, we expect the quantum number of standing probability wave that will fit this distance to increase. We also expect that for a given velocity, the wavelength does not change. Similarly, as the velocity, $\vpar$, increases, the wavelength will decrease and so we expect the quantum number of standing probability wave that matches this velocity to increase as well. Those are precisely the patterns observed in Figs.\ 9, 10 and 11. One discrepancy between the model and the numerical results can be noted: For $d$ = $9a_s$ and $13a_s$ some bands are intermittently missing. This is likely due to forbidden gaps in the subband structure that exist for these vicinal widths within the $\vpar$ range considered which, however, is the part of finer details of the band structure. Furthermore, note that the resonance peak positions for the two different strikes quickly merge into the same graph for $d$ = 5 but they merge more slowly as $d$ increases. In fact, for $d$ = 5 the step strike and midpoint strike peaks occur at the same points, whereas for $d$ = 13 they occur at the same points only for higher values of $\vpar$.  This is also the trend the model curves indicate. 

We may further note that owing to the slightly different image state energies $E$ among the substrate surfaces considered (see Fig.\ 7), the model \eq{sq-well} plots slightly different curves from one metal to the other even for a given image state. As already pointed out in subsection IIIA, this is likely the prime reason why the peak positions in ion survival for a given step size $d$ suffers a small offset from one vicinal surface to the other. Further, small differences between the peak positions for the choice of BZ versus BSL trajectory for a given vicinal surface, as too noted in subsection IIIA, is clearly evident by noting the positions of the symbols with respect to the model lines which are independent of a trajectory.

\section{Conclusion}

Using the metals Cu(111), Au(100) and Pd(111) vicinally miscut in a selection of terrace sizes within a few nanometers we simulate the dynamics of the active-electron in a hydrogen anion projectile impinging at shallow angles to the surface.  To do so, we used a fully quantum mechanical wave packet propagation methodology. In the absence of a completely ab\,initio potential, we used a simple but successful model of vicinal stepping. The electron dynamics is visualized by animating the wave packet probability density in real time. The results show structures in the ion survival probability due to the image state subband dispersion introduced by the vicinal superlattice. This produced structures in the ion survival as a function of terrace sizes, proving the ability of our propagation methodology to study RCT tunneling between anions and surfaces with superperiodicity in the nanoscale range. Detailed differences in the RCT dynamics in general and the subband resonance signals in particular among choices of substrate surfaces are motivated by the interference between substrate and vicinal step dispersions. Also, the role of ion-surface effective time of interaction by selecting two different classical trajectories are unveiled. Even though the calculation is based on one-active-electron model, effects of electron correlations due to occupied valence band can only enhance the recapture rate by impeding decay into the bulk from the Pauli blockade. The effects uncovered can be observed in grazing ion spectroscopy within the current laboratory technology, although accessing effects of the azimuthal orientation of the scattering plane will require full 3D simulation. The invariance of the results to the strike location on the surface over the higher parallel velocity range (results not shown here, but was demonstrated previously~\cite{shaw18}) suggests that for a sufficiently grazing flyby aiming the ion beam is likely irrelevant, providing some experimental freedom. 

To this end, our results indicate that one interesting spectroscopic pathway to probe subband dispersions in vicinal superlattice is RCT studies of negative ions in grazing scattering off stepped metal surfaces. As we show, the simpler approach to that goal is to choose the impact energies and scattering angles in such a way that the ion velocity perpendicular to the flat surface direction stays constant while varying in the parallel direction. This will keep the interaction largely free of alterations from the band structure perpendicular to the precursor flat surface. It is true that the results in Figs.\ 4, 5, and 6 show variations in the range of $\pm$0.2\% to $\pm$0.4\% around a small average H$^-$ survival probability (ion fraction) of roughly 2-5\%. However, experimentally an ion fraction as little as 0.1\% has been measured for flat surfaces within an error range of about $\pm$0.1\%~\cite{guillemot} over similar ion-velocities used in this work. Therefore, we believe that structures in our current predictions should be observable in the experiment.

\begin{acknowledgments}
The authors would like to acknowledge the support received by an Extreme Science and Engineering Discovery Environment (XSEDE) allocation Grant for high performance computation, which is supported by National Science Foundation (NSF) grant number ACI-1053575. The research was also supported in part by NSF Grant Number PHY-1806206.
\end{acknowledgments}

\bibliographystyle{apsrev4-1}

\begin{thebibliography}{99}
\bibitem{pratt} 
S.~J. Pratt and S.~J. Jenkins, Beyond the surface atlas: A roadmap and gazetteer for surface symmetry and structure, Surf.\ Sci.\ Rep. \textbf{62}, 373 (2007).

\bibitem{tegencamp09} C.~Tegenkamp, Vicinal surfaces for functional nanostructures, J. Phys.: Condens.\ Matter \textbf{21}, 013002 (2009).

\bibitem{schiller05} F.~Schiller, M.~Ruiz-Os\'{e}s, J.~Cord\'{0}n, and J.~E. Ortega, Scattering of Surface States at Step Edges in Nanostripe Arrays, \prl\, \textbf{95}, 066805 (2005).

\bibitem{didiot}
C.~Didiot, S.~Pons, B.~Kierren, Y.~Fagot-Revurat and D.~Malterre, Nanopatterning the electronic properties of gold surfaces with self-organized superlattices of metallic nanostructures, Nature Nanotech. \textbf{2}, 617 (2007).
 
\bibitem{suzuki}
K.~Suzuki, K.~Kanisawa, C.~Janer, S.~Perraud, K.~Takashina, T.~Fujisawa and Y.~Hirayama, Spatial imaging of two-dimensional electronic states in semiconductor quantum wells, \prl\, \textbf{98}, 136802 (2007).
 
\bibitem{bolz}
A.~Bolz, C.~Meyer, C.~Heyn, W.~Hansen, M.~Morgenstern and R.~Wiesendanger, Wave-function mapping of InAs quantum dots by scanning tunneling spectroscopy, \prl\, \textbf{98}, 196804 (2007).

\bibitem{mugarza}
A.~Mugarza and J.~E. Ortega, Electronic states at vicinal surfaces, J. Phys.: Condens.\ Matter \textbf{15}, S3281 (2003).

\bibitem{mugarza01} A.~Mugarza, A.~Mascaraque, V.~Pérez-Dieste, V.~Repain, S.~Rousset, F.~J. Garc\'{i}a de Abajo, and J.~E. Ortega, \prl\, \textbf{87}, 107601 (2001).

\bibitem{ortega}
J.~E. Ortega, S.~Speller, A.~R. Bachmann, A.~Mascaraque, E.~G. Michel, A.~Narmann, A.~Mugarza, A.~Rubio and F.~J. Himpsel, Electronic wave function at vicinal surfaces: Switch from terrace to step modulation, \prl\, \textbf{84}, 6110 (2000).
 
\bibitem{baumberger}
F.~Baumberger, T.~Greber and J.~Osterwalder, Fermi surfaces of the two-dimensional surface states on vicinal Cu(111), \prb\, \textbf{64}, 195411 (2001).
 
\bibitem{greber}
F.~Baumberger, T.~Greber and J.~Osterwalder, Step-induced one-dimensional surface state on Cu(332), \prb\, \textbf{62}, 15431, (2000).
 
\bibitem{lei}
J.~Lei, H.~Sun, K.~W. Yu, S.G.~Louie and M.~L. Cohen, Image potential states on periodically corrugated metal surfaces, \prb\, \textbf{63}, 045408 (2001).

\bibitem{shaw18}
J. Shaw, D. Monismith, Y. Zhang, D. Doerr, and H. S. Chakraborty, Ion Survival in grazing collisions of H$^-$ with vicinal nanosurfaces as a probe for subband electronic structures, \pra\, \textbf{98}, 052705 (2018).

\bibitem{rabalais03} 
J.W. Rabalais, {\em Principles and Applications of Ion Scattering Spectrometry: Surface Chemical and Structural Analysis}, 
        (Wiley-Interscience, Hoboken, New Jersey, 2003).

\bibitem{stout00} 
K.J. Stout and L. Blunt, {\em Three-Dimensional Surface Topography}, (Penton Press, London, 2000).

\bibitem{korkin07} 
{\em Nanotechnology for Electronic Materials and Devices}, edited by A. Korkin, J. Labanowski, E. Gusev, and S. Luryi, (Springer, New York, 2007).

\bibitem{campbell01} 
S.A. Campbell, {\em The Science and Engineering of Microelectronic Fabrication}, (Oxford University Press, New York, 2001).

\bibitem{canario03} 
A.~R. Canario, E.~A. Sanchez, Yu. Bandurin, and V.~A. Esaulov, Growth of Ag nanostructures on TiO$_2$(110), Surf.\ Sci.\ Lett. \textbf{547}, L887 (2003).
 
\bibitem{bahrim}
B.~Bahrim, B.~Makarenko and J.~W. Rabalais, Band gap effect on H$^-$ ion survival near Cu surfaces, Surf.\ Sci.\ \textbf{594}, 62 (2005).
 
\bibitem{yang}
Y.~Yang and J.~A. Yarmoff, Charge exchange in Li scattering in Si surfaces, \prl\, \textbf{89}, 196102 (2002).
 
\bibitem{hecht}
T.~Hecht, H.~Winter, A.~G. Borisov, J.~P. Gauyacq and A.~K. Kazansky, Role of the 2D surface state continuum and projected band gap in charge transfer in front of Cu(111) surface, \prl\, \textbf{84}, 2517 (2000).
 
\bibitem{guillemot}
L.~Guillemot and V.~A. Esaulov, Interaction time dependence of electron tunneling processes between an atom and a surface, \prl\, \textbf{82}, 4552 (1999).
 
\bibitem{sanchez} 
E.~Sanchez, L.~Guillemot and V.~A. Esaulov, Electron transfer in the interaction of Flourine and Hydrogen with Pd(100), \prl\, \textbf{83}, 428 (1999).

\bibitem{gainullin15} 
I.~K. Gainullin and M.~A. Sonkin, Three-dimensional effects in resonant charge transfer between atomic particles and nanosystems, \pra\, \textbf{92}, 022710 (2015).

\bibitem{hakala07} 
T.~Hakala, M.~J. Puska, A.~G. Borisov, V.~M. Silkin, N.~Zabala, and E.~V. Chulkov, Excited states of Na nanoislands on the Cu(111) surface, \prb\, \textbf{75}, 165419 (2007).

\bibitem{obreshkov06} B.~Obreshkov and U.~Thumm, Neutralization of H$^-$ near vicinal metal surfaces, \pra\, \textbf{74}, 012901 (2006).
 
\bibitem{chak70}
H.~S. Chakraborty, T.~Niederhausen and U.~Thumm, Resonant neutralization of H$^-$ near Cu surfaces: Effects of the surface symmetry and ion trajectory, \pra\, \textbf{70}, 052903 (2004).
 
\bibitem{chak69}
H.~S. Chakraborty, T.~Niederhausen and U.~Thumm, Effects of the surface Miller index on the resonant neutralization of hydrogen anions near Ag surfaces, \pra\, \textbf{69}, 052901 (2004).
 
\bibitem{chak241}
H.~S. Chakraborty, T.~Niederhausen and U.~Thumm, On the effect of image states on resonant neutralization of hydrogen anions near metal surfaces, Nucl.\ Instrum.\ Methods B \textbf{241}, 43 (2005).
 
\bibitem{schmitz}
A.~Schmitz, J.~Shaw, H.~S. Chakraborty and U.~Thumm, Band-gap-confinement and image-state-recapture effects in the survival of anions scattered from metal surfaces, \pra\, \textbf{81}, 042901 (2010).

\bibitem{swamy99}
K.~Swamy, E.~Bertel, and I.~Vilfan, Step interaction and relaxation at steps: Pt(110), Surf.\ Sci.\ \textbf{425}, 369 (1999).
 
\bibitem{chulkov}
E.~V. Chulkov, V.~M. Silkin and P.~M. Echenique, Image potential states of metal surfaces: Binding energies and wave functions, Surf.\ Sci. \textbf{437}, 330 (1999).

\bibitem{chulkov97}
E.~V. Chulkov, V.~M. Silkin and P.~M. Echenique, Image potential states on lithium, copper and silver surface, Surf.\ Sci. \textbf{391}, L1217 (1997).
 
\bibitem{ermoshin}
V.~A. Ermoshin and A.~K. Kazansky, Wave packet study of H$^-$ decay in front of a metal surface, Phys.\ Letts. A \textbf{218}, 99 (1996).
 
\bibitem{press}W.~H. Press, S.~A. Teukolsky, W.~T. Vetterling and B.~P. Flannery, \textit{Numerical Recipes: The Art of Scientific Computing} (Cambridge University Press, Cambridge, 2007).
 
\bibitem{biersack} 
J.~P. Biersack and J.~F. Ziegler, Refined universal potentials in atomic collisions, Nucl.\ Instrum.\ Methods \textbf{194}, 93 (1982).

\bibitem{borisov96} 
A.~G. Borisov, D.~Teillet-Billy, J.~P. Gauyacq, H.~Winter, and G.~Dierkes, Resonant charge transfer in grazing scattering of alkali-metal ions from an Al(111) surface, \prb\, \textbf{54}, 17166 (1996).

\bibitem{mueller70}
F.~M. Mueller, A.~J. Freemant, J.~O. Dimmock, and A.~M. Furdyna, Electronic structure of palladium, \prb\, \textbf{1}, 4617 (1970).

\bibitem{winter91}
H. Winter, Charge transfer in grazing ion-surface scattering, Commts.\ At.\ Mol.\ Phys. \textbf{26}, 287 (1970).

\bibitem{osma}
J. Osma, I. Sarria, E.~V. Chulkov, J.~M. Pitarke, and P.~M. Echenique, Role of the intrinsic surface state in the decay of image states at a metal surface, \prb\, \textbf{59}, 10591 (1999).

\bibitem{gao}
S. Gao, and D.~C. Langreth, Image state mediated electron transfer at surfaces, Surf.\ Sci.\ Letts., \textbf{398}, L314 (1998).

\bibitem{chulkov98}
E.~V. Chulkov, I. Sarria, V.~M. Silkin, J.~M. Pitarke, and P.~M. Echenique, Lifetimes of Image-Potential States On Copper Surfaces, \prl\, \textbf{80}, 4947 (1998).

\end{thebibliography}

\end{document}